\documentclass[runningheads]{llncs}

\usepackage{graphicx} 
\renewcommand{\baselinestretch}{1} 
\usepackage[margin=1.2in]{geometry}
\usepackage[style=nature,citestyle=numeric,sorting=none,sortcites=true,autolang=hyphen,backref=false,backend=biber,defernumbers=false, hyperref=false, maxbibnames=99]{biblatex}
\usepackage{appendix}
\usepackage{helvet}
\usepackage[table,xcdraw]{xcolor}
\usepackage{longtable}
\usepackage{pdfpages}

\usepackage{tikz}
\usepackage{amsmath}
\usepackage[nottoc]{tocbibind}
\usepackage{etoolbox}
\title{Rectified flow-based prediction of post-treatment brain MRI from pre-radiotherapy priors for patients with glioma}
\author{Selena Huisman \inst{1,2}, Nordin Belkacemi \inst{1,2}, Vera C. Keil \inst{2,3,4}, Joost J.C. Verhoeff \inst{1,2}, Szabolcs David \inst{1,2}\\
Email corresponding author: s.i.huisman@amsterdamumc.nl}
\titlerunning{ }
\authorrunning{ }
\institute{Amsterdam UMC, Department of Radiation Oncology, \\
De Boelelaan 1117, 1081 HV Amsterdam, The Netherlands \inst{1} \\
Cancer Center Amsterdam, Imaging and Biomarkers, \\
De Boelelaan 1117, 1081 HV Amsterdam, The Netherlands \inst{2}\\
Department of Radiology and Nuclear Medicine, \\
De Boelelaan 1117, 1081 HV Amsterdam, The Netherlands \inst{3}\\
Amsterdam UMC, Vrije Universiteit Amsterdam, Amsterdam Neuroscience, Brain Imaging \\
De Boelelaan 1117, 1081 HV Amsterdam, the Netherlands \inst{4}}

\date{June 2025}

\begin{document}
\maketitle
\begin{abstract}
    \textbf{Purpose/Objective:} Brain tumors result in 20 years of lost life on average. Standard therapies induce complex structural changes in the brain that are monitored through MRI. Recent developments in artificial intelligence (AI) enable conditional multimodal image generation from clinical data. In this study, we investigate AI-driven generation of follow-up MRI in patients with intracranial tumors through conditional image generation. This approach enables realistic modeling of post-radiotherapy changes, allowing for treatment optimization.

    \textbf{Material/Methods:} The public SAILOR dataset of 25 patients was used to create a 2D rectified flow model conditioned on axial slices of pre-treatment MRI and RT dose maps. Cross-attention conditioning was used to incorporate temporal and chemotherapy data. The resulting images were validated with structural similarity index measure (SSIM), peak signal-to-noise ratio (PSNR), Dice scores and Jacobian determinants.
    
    \textbf{Results:} The resulting model generates realistic follow-up MRI for any time point, while integrating  treatment information. Comparing real versus predicted images, SSIM is 0.88, and PSNR is 22.82. Tissue segmentations from real versus predicted MRI result in a mean Dice-Sørensen coefficient (DSC) of 0.91. The rectified flow (RF) model enables up to 250x faster inference than Denoising Diffusion Probabilistic Models (DDPM).
    
    \textbf{Conclusion:} The proposed model generates realistic follow-up MRI in real-time, preserving both semantic and visual fidelity as confirmed by image quality metrics and tissue segmentations. Conditional generation allows counterfactual simulations by varying treatment parameters, producing predicted morphological changes. This capability has potential to support adaptive treatment dose planning and personalized outcome prediction for patients with intracranial tumors. Code will be available upon peer-reviewed publication at: https://github.com/SelenaIHuisman/RF-GlioPREDICT

\end{abstract}
\clearpage

\setcounter{page}{1}
\section{Introduction}

Brain tumors result in one of the highest average years of life lost \cite{yearsoflostlife} ($>$20 years) among all cancers, contributing to a significant societal \cite{CNSyearsoflostlife} and economic burden \cite{USAtumors}. Approximately 50-90\% of patients live sufficiently long ($>$6 months) to experience irreversible, progressive cognitive decline \cite{neurocognitivefunctioning}, often manifesting in multiple cognitive domains, including memory, attention, and executive function \cite{cognitivedisability}. 

Standard of care for brain tumors consists of a combination of therapies, including surgery, radiation therapy (RT), and systemic therapies such as immunotherapy and chemotherapy \cite{EANOguideline}. The variability in outcome between patients is high and arises from heterogeneous tumor biology \cite{tumorbiology}, patient-specific anatomy \cite{patientanatomy}, and treatment context \cite{treatmentcontext}, which interact in a non-linear manner over time. Treatment response manifests as complex morphological changes affecting both diseased and healthy brain tissue, which are monitored through standard magnetic resonance imaging (MRI), commonly including T1-weighted, T1-weighted with gadolinium contrast enhancement (T1Gd) and T2 fluid attenuated inversion recovery (T2-FLAIR) weighter pulse sequences. \cite{volumeloss,corticalthickness,greymattervolumeloss}. 

Each of these modalities provides complementary contrast of both brain tissue and tumor, enabling quantitative assessment of treatment-related tissue changes, tumor burden, and disease progression over time. A commonly used method to assess treatment response on MRI is the RANO criteria, which can influence treatment decisions on continuation, adaptation or discontinuation of therapy \cite{RANO2}. Owing to these properties, routine clinical MRI is essential for monitoring disease progression and treatment response. 
 Therefore, the ability to accurately predict post-treatment MRI findings before treatment is applied could enable precise outcome prediction, earlier treatment adaptation and personalized care strategies.

Recent advances in artificial intelligence (AI) such as conditioned denoising diffusion probabilistic models (DDPMs) \cite{ho2020denoising}, allow for realistic image generation from multimodal data, such as Imagic \cite{Imagic} for image editing or Gemma 3 \cite{Gemma3} and GPT 4o \cite{GPT4o} for multimodal image generation. AI-based image generation is already applied in various medical applications, such as image registration \cite{voxelmorph1,voxelmorph2, synthmorph}, fast MRI reconstruction \cite{fastMRI} and segmentation \cite{TADiff, SynthSeg, SynthSegRobust}. Furthermore, image generation models are already transforming radiotherapy by enabling synthetic CT generation from MRI for treatment planning, making MRI-only planning workflows increasingly available in clinical practice \cite{SyntheticCT}.

Cross-attention mechanisms \cite{vaswani2023attention} and timestep embeddings \cite{TADiff} enable diffusion-style models to supplement the imaging data learned by the denoising \cite{unetarchitecture2016} with textual or numerical information such as clinical variables and temporal data. Treatment-aware diffusion (TADiff) utilizes DDPM to predict late follow-up MRIs from pre-RT MRIs, dose and early follow-up MRIs and timestep embeddings for chemotherapy and temporal information \cite{TADiff}. Another example by Pinaya et al. uses latent diffusion models to generate 3D brain MRI from noise, conditioning on age, gender, ventricular volume, and brain volume via both concat- and cross-attention \cite{Pinaya2022brainLDM}. These emerging methods show that image generation is feasible for brain MRI.

Most existing longitudinal generative models for brain MRI focus on non-modifiable drivers such as aging or neurodegenerative disease progression, where conditioning variables like age, baseline anatomy or disease stage cannot be altered at the individual-patient level \cite{healthypred,longalz}. Even when treatment is included, as in treatment-aware diffusion approaches that encode chemotherapy and time since treatment \cite{TADiff}, it is typically modeled as a global categorical or scalar variable rather than as a spatially resolved prescription. In contrast, radiotherapy offers a fundamentally different use case: the conditioning variables of interest are modifiable treatment decisions, which are three dimensional and high resolution for dose. This opens up the possibility of using generative models to explore alternative radiotherapy strategies by changing modifiable inputs such as dose distribution or fractionation scheme, and to predict how these different, yet clinically viable treatments would affect the brain. With these generated images, clinicians could adjust RT dose and dose distribution to minimize effects in healthy-appearing brain tissue while maximizing the effect on the tumor regions.

Adjustments to the RT dose can only be applied to the patient if, ideally, all variables in the models are known before treatment. Current state-of-the-art models require multiple longitudinal MRIs \cite{TADiff}, therefore counterfactual generations cannot affect treatment decisions, as they are not available before RT treatment is applied. Hence, a model that includes exclusively pre-treatment MRIs, RT dose maps, and other treatment information would enable clinicians to use the generated images as a basis for treatment decisions.

To enable flexible, clinically usable applications of MRI prediction, inference speed is critical. Although the image quality of the initial diffusion-style models is high, their sampling speed is limited by the number of sampling steps required to denoise an image. Recent research has focused on improving these sampling speeds by either solving the DDPM ordinary differential equation (ODE) more quickly or by changing the ODE to require fewer sampling steps. Major improvements in sampling speed have been achieved with techniques such as DPM-solver \cite{DDPMsolver}, diffusion denoising implicit models (DDIM) \cite{DDIM2022}, and, most recently, rectified flow (RF) \cite{rflow}. DDPM required an average of 1000-4000 denoising steps \cite{Nichol2021improveddiff, ho2020denoising}, DDIM was able to generate images in 50-250 steps \cite{Nichol2021improveddiff, ho2020denoising}, while RF can produce even faster inference speeds of 1-30 denoising steps \cite{rflow, maisi} by straightening the paths from noise to prediction. Hence, rectified flow-based models could enable real-time prediction of brain MRI.

This study investigates real-time AI-driven generation of follow-up MRIs in patients with brain tumors using rectified flow–based conditional image generation. The proposed model generates synthetic assumed follow-up MRI outcomes from pre-radiotherapy MRIs, temporal information, radiotherapy dose distributions and chemotherapy timing. To achieve this, we combine rectified flow with cross-attention and concat-based conditioning, enabling real-time counterfactual MRI generation driven by modifiable, spatially resolved treatment inputs. The aim of such predictions is to provide clinicians with a novel tool for personalized treatment planning through improved outcome prediction.

\section{Methods}

\subsection{Data}

 The freely available, public SAILOR dataset \cite{SAILOR} was used for this study. The dataset was created from patients treated at Oslo University Hospital and consists of 27 high grade glioma cases, each of which have a planning MRI and 3 to 19 follow-up MRIs. Each patient was treated with the standard of care Stupp protocol \cite{StuppProtocol} including debulking surgery, followed by chemoradiotherapy (CRT) with concomitant temozolomide (TMZ) and six cycles of adjuvant TMZ chemotherapy. Two patients were excluded due to issues with the RT dose map, resulting in a dataset of 25 patients. Median age is 56 years, with 8 females and 19 males. In total, there are 257 MRI timepoints, of which 25 are at baseline (before RT treatment), with an average of 10 MRI scans per patient. The training set was created with a randomly sampled set of 21 patients, with 2 patients each for validation and testing, respectively.
 
\subsection{Preprocessing}

The dataset comes preprocessed with the following steps: defacing, N4-based bias-field correction, ANT-based denoising, FSL-based skull stripping, rigid intra-patient registration and affine registration to the MNI brain template, as well as longitudinal intensity normalization \cite{SAILOR}. In addition to the preprocessing already applied, the 3D images were normalized between 0 and 1 and scaled to 0.5\% and 99.5\%. The RT dose maps were rescaled between 0 and 1, with 1 being the highest dose in the patient population; the relative dose among patients is preserved. Finally, the 3D images were presliced into axial slices between slice 42 and slice 140, which covers the majority of the supratentorial part of the brain. Each slice was labeled either healthy-appearing or diseased based on the ONCOhabitats segmentation \cite{ONCOhabitats} provided along with the dataset. During training, slices were sampled with a balanced data sampler, resulting in a fifty-fifty split of healthy-appearing and diseased slices. 

\subsection{Model architecture and hyperparameters}

 MONAI was used for model construction, data loading and augmentations \cite{monai}. In total, four models were trained, with the following combinations of conditioning variables: no treatment conditioning, only RT dose conditioning, only chemotherapy conditioning and both chemotherapy and dose conditioning. Time and pre-treatment MRIs were always included as conditioning. All models were trained with a batch size of 8 with a gradient accumulation of 2, on a single Nvidia H100 GPU. Each model took around 30 hours to train. Models were trained for 140-160 epochs, where the best model was selected based on the lowest validation loss.

\begin{figure}
     \centering
     \includegraphics[width=1\linewidth]{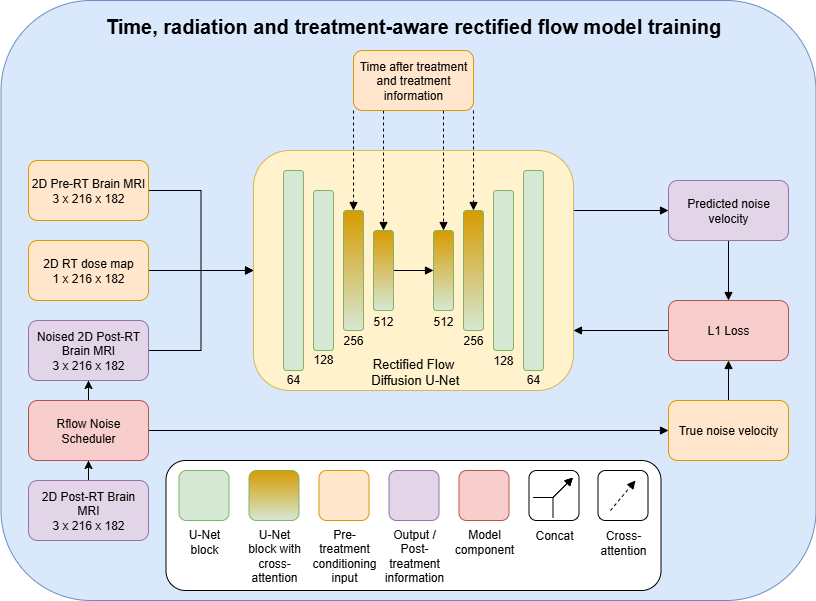}
     \caption{The architecture and training process of the rectified flow diffusion U-Net. The model is trained with pre-treatment information as conditioning, and post treatment information for the L1 loss function.}
     \label{fig:architecture}
 \end{figure}

 Model architecture is shown in Figure \ref{fig:architecture}. Noise is added to 2D 216x184 follow-up MRIs in 3 modalities (T1, T1Gd, T2-FLAIR), which are then concatenated with the spatial conditioning of 2D baseline MRI and 2D RT dose maps. Vector-based conditioning is added via cross-attention blocks in the 3\textsuperscript{rd} and 4\textsuperscript{th} layers of the model. MAE loss is computed betwen the learned velocity of the noise and its true velocity of the noise, then backpropagated to train the model.

\begin{figure}
     \centering
     \includegraphics[width=1\linewidth]{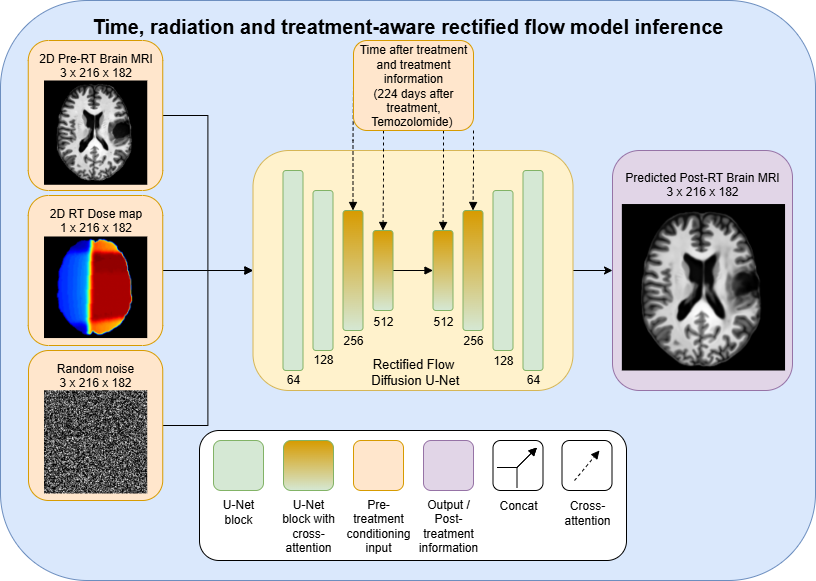}
     \caption{The architecture and inference process of the rectified flow diffusion U-Net. Follow-up MRIs are generated for any timepoint and treatment combination based on the input information.}
     \label{fig:inference}
 \end{figure}

 Figure \ref{fig:inference} shows the inference process. The 2D baseline MRI and RT dose maps are concatenated with noise in the shape of the follow-up images and fed into the model. The model then predicts the follow-up images based on the number of days after baseline, type of chemotherapy, RT dose map and baseline MRI.
 
\clearpage

\subsection{Validation}

All four models were validated on the test set using Peak Signal to Noise Ratio (PSNR), Structural Similarity Index Measure (SSIM) and Mean Squared Error (MSE). Only the model with both RT and chemotherapy conditioning was validated using Dice scores (DSC) of tissue and cerebrospinal fluid and Jacobian determinants \cite{DBM1, DBM2}. One test subject received surgery after radiotherapy, and the follow-ups after the operation were excluded, since the model is not trained to predict whether surgery has taken place. All other follow-up MRIs for these patients were included in the test set.

PSNR, SSIM and MSE were calculated for each slice of the test set. Additionally, each slice was segmented with FSL's FAST \cite{FSL}. From these segmentations, DSC and volumes were calculated for each slice using two categories: tissue and CSF. Finally, we used FNIRT \cite{FSL} non-linear registration to estimate local contraction and expansion via log Jacobian determinants between the baseline and real follow-up and between the baseline and predicted follow-up for all slices. Mean absolute Jacobian determinants were calculated for tissue and CSF together to show the total effect size, while mean Jacobian determinants were used for tissue and CSF separately to show the direction of the effect. Background voxels were excluded from the analysis. Finally, the model was sampled for different treatment combinations and predictions over time, creating counterfactual simulations for demonstration purposes.

\section{Results}

\begin{figure}
    \centering
    \includegraphics[width=1\linewidth]{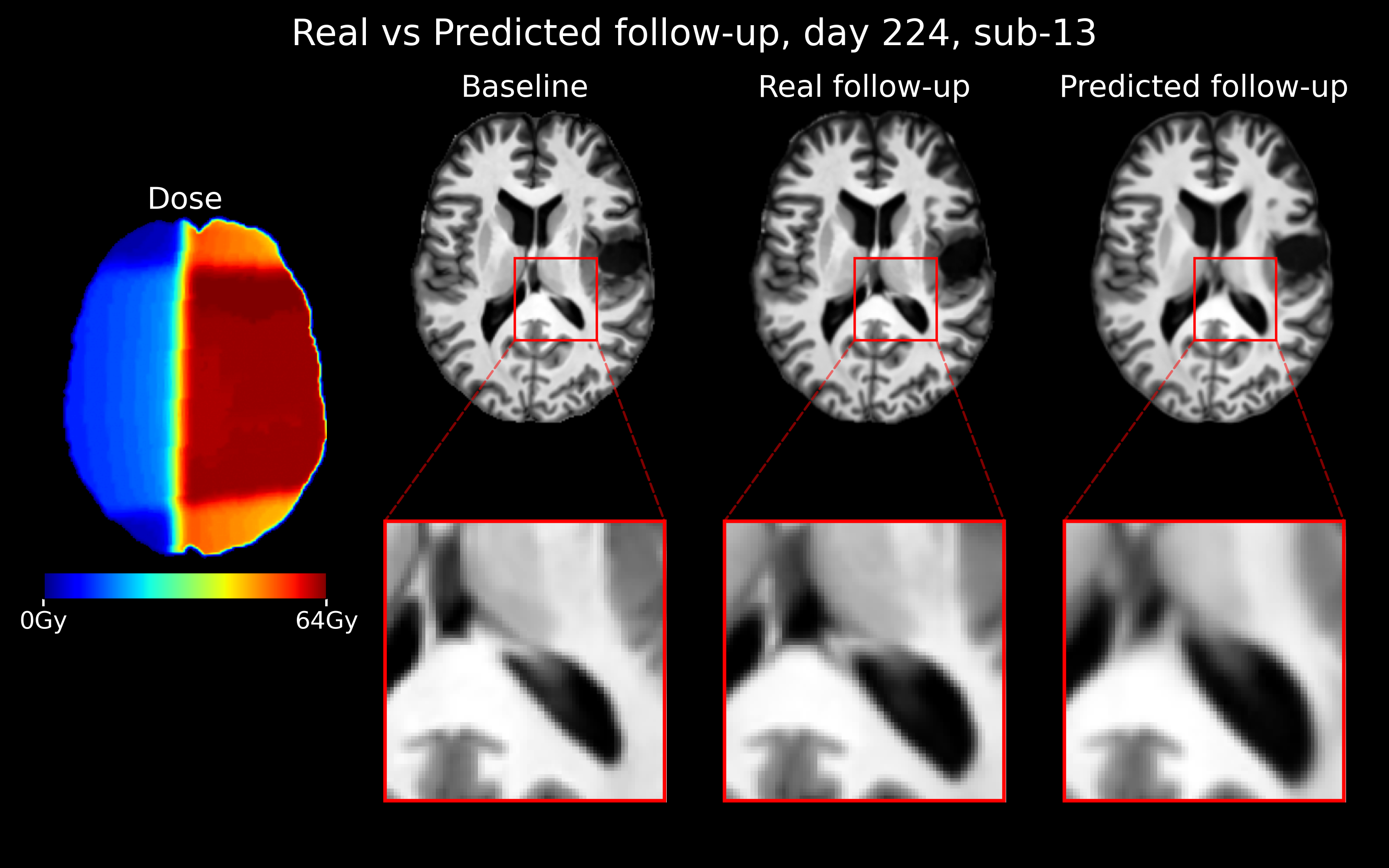}
    \caption{An example slice from the middle of the brain, with the radiotherapy dose, real baseline MRI, real follow-up MRI and a predicted follow-up MRI. Zoomed-in sections of the ventricles are shown below the images to highlight local morphological changes predicted by the model. For the RT dose, warmer colors mean a higher dose. For each image, right is on the right.}
    \label{fig:example}
\end{figure}

The proposed RT- and chemotherapy-conditioned model generates realistic post-treatment MRIs, as shown in Figure \ref{fig:example}. Visually, the ventricles in the predicted follow-up are similar to the real follow-up. Quantitatively, the images are similar, with an SSIM of 0.88 (SD: 0.08), MSE of 0.008 (SD: 0.007), and PSNR of 22.82 (SD: 4.02). The model with both conditioning methods does not achieve the best performance of the tested models, since the model with no treatment conditioning achieves an SSIM of 0.89, while the model with only chemotherapy conditioning achieves a PSNR of 23.67 and MSE of 0.006. Further quantitative results for models with different combinations of treatment conditioning are available in Supplementary Table 1.

DSC for CSF and tissue are 0.83 (SD: 0.13) and 0.91 (SD: 0.07), respectively. CSF volume was on average 5717 mm\textsuperscript{3} (SD: 1320 mm\textsuperscript{3}) for the real follow-ups and 5893 mm\textsuperscript{3} (SD: 1418 mm\textsuperscript{3}) for the predicted follow-ups, indicating significantly more CSF in the predicted images compared to the real images (P = 0.0007). Tissue volume for real follow-ups was 11165 mm\textsuperscript{3} (SD: 2529 mm\textsuperscript{3}) while predicted follow-up volume was 11008 mm\textsuperscript{3} (SD: 2421 mm\textsuperscript{3}) on average, indicating that real follow-ups contain slightly more tissue on average compared to predicted follow-ups. However, the difference in tissue volume is not significant (P = 0.0951).

\clearpage

\begin{figure}[ht]
    \centering
    \includegraphics[width=0.92\linewidth]{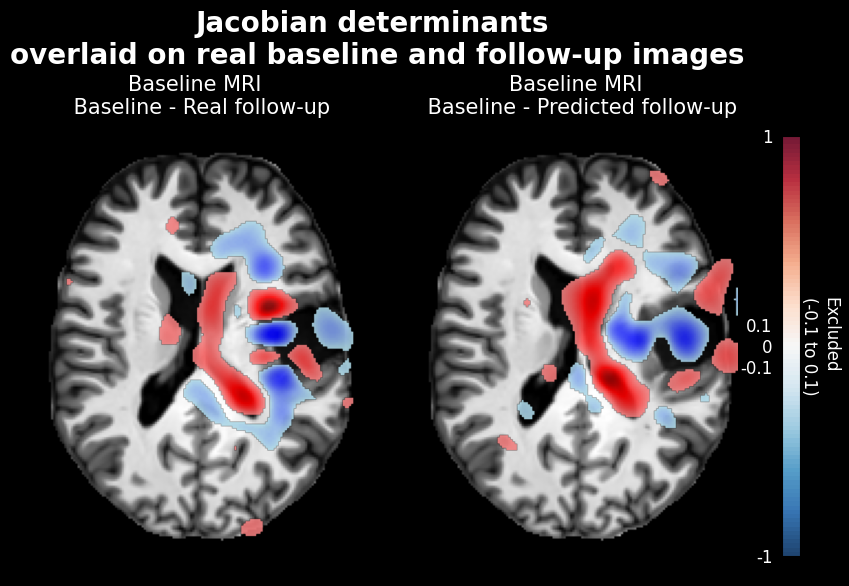}
    \caption{The Jacobian determinants between baseline and real follow-up and baseline and predicted follow-up, with the baseline image underlaid. Red means that a pixel expands during non-linear registration, while blue means that the pixel contracts. Values between -0.1 and 0.1 are excluded. For each image, right is on the right.}
    \label{fig:Jacobian}
\end{figure}

The log Jacobian determinants, as shown in Figure \ref{fig:Jacobian}, indicate that there is a similar amount of absolute change in tissue and CSF from baseline to real follow-up (mean: 0.069, SD: 0.027) as from baseline to predicted follow-up (mean: 0.067, SD: 0.023). When analyzing tissue and CSF separately, CSF shows more contraction in predicted follow-ups (mean: -0.024) compared to real follow-ups (mean: -0.0096). Changes in tissue are similar between real and predicted, with a mean of -0.0092 for real images and -0.011 for predicted follow-ups.

\begin{figure}[ht]
    \centering
    \includegraphics[width=0.75\linewidth]{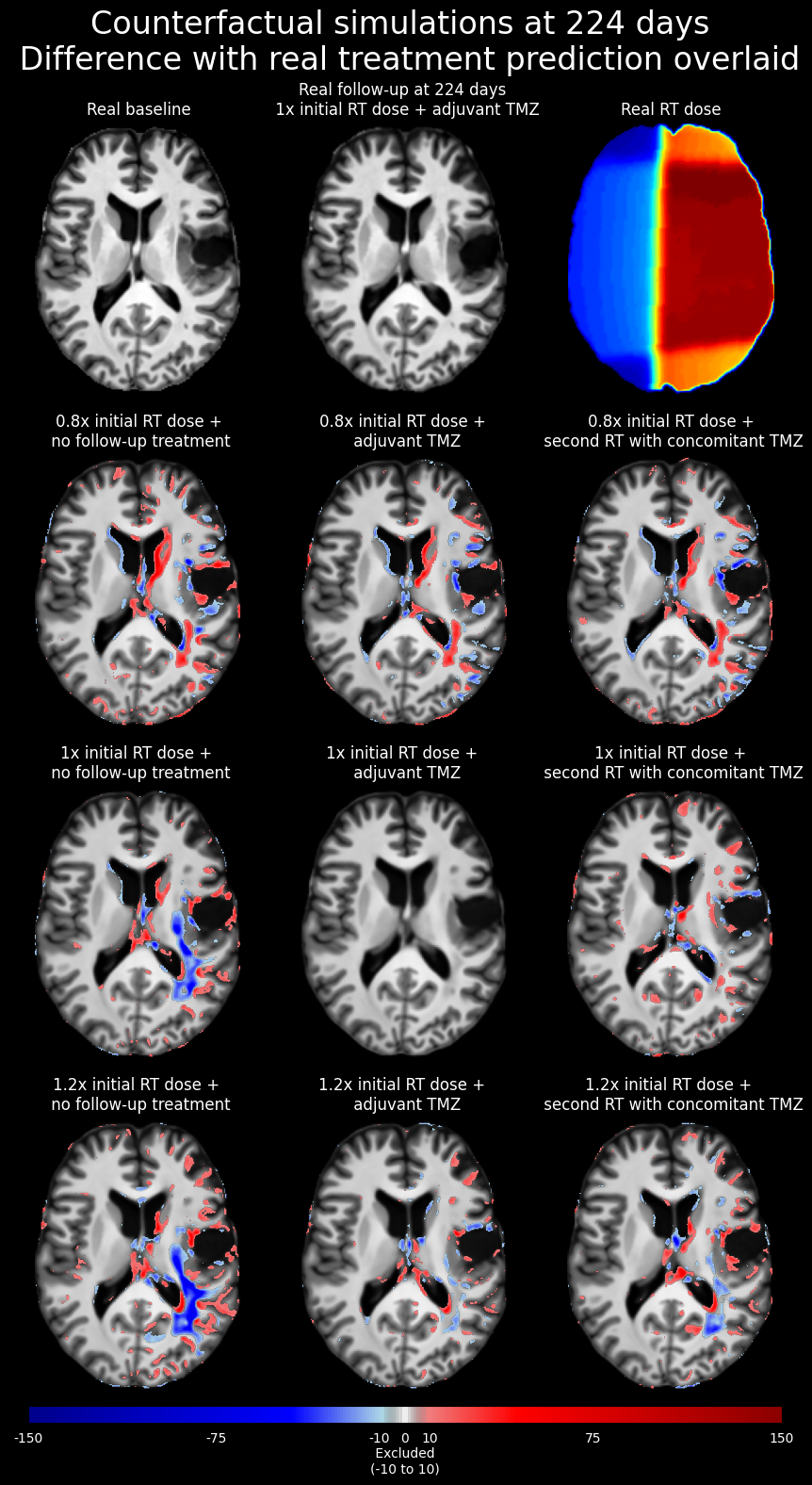}
    \caption{Counterfactual simulations showing how different treatment combinations affect the brain. The baseline, real followup and the true treatment (1x RT dose and temozolomide (TMZ) are shown in the top row as reference. The bottom 3 x 3 MRI slices show the prediction outcomes at 224 days, each for a different treatment combination. The predicted images are all subtracted from the predicted image based on the true treatment (the middle one), resulting in an overlay that shows pixel-intensity differences. Red means positive values, while blue means negative values. The baseline, real follow-up and RT dose are exactly the same as figure \ref{fig:example}. For each image, right means right. Values between -10 and 10 are excluded for clarity.}
    \label{fig:treatment_counterfactuals}
\end{figure}

Figure \ref{fig:treatment_counterfactuals} shows counterfactual simulations based on changes to treatment for a patient, predicting for the same timepoint to demonstrate that different treatments lead to different (radiological) outcomes. The initial RT dose was either not adjusted, multiplied by 0.8 or multiplied by 1.2 in every voxel. For the timepoint generated (day 224), either no treatment was simulated, adjuvant temozolomide or a second round of RT with concomitant temozolomide. For the real timepoint, TMZ and 1x (true prescribed) radiotherapy dose were given to the patient. The center image represents the predicted image for this treatment, while the surrounding images show the predicted images for counterfactual treatments. Generated images laid out in the cardinal directions compared to the center represent a change in a single treatment, while images in the intermediate directions have both treatment variables adjusted. Reduction in the pixel-intensity of brain tissue (blue) around the ventricles generally indicates an increase in ventricular size, indicating tissue loss. Conversely, red in the ventricles indicates an increase in tissue. In tissue, blue indicates swelling or edema since there is a reduction of intensity. An increase in intensity in locations with edema would indicate that the edema decreases; for this patient, less RT dose means less tissue atrophy (red around the ventricles), but not treating with chemotherapy results in a prediction of edema (blue) in the right side of the brain.

\clearpage

\clearpage

\begin{figure}[ht]
    \makebox[\textwidth][c]{\includegraphics[width=1.2\linewidth]{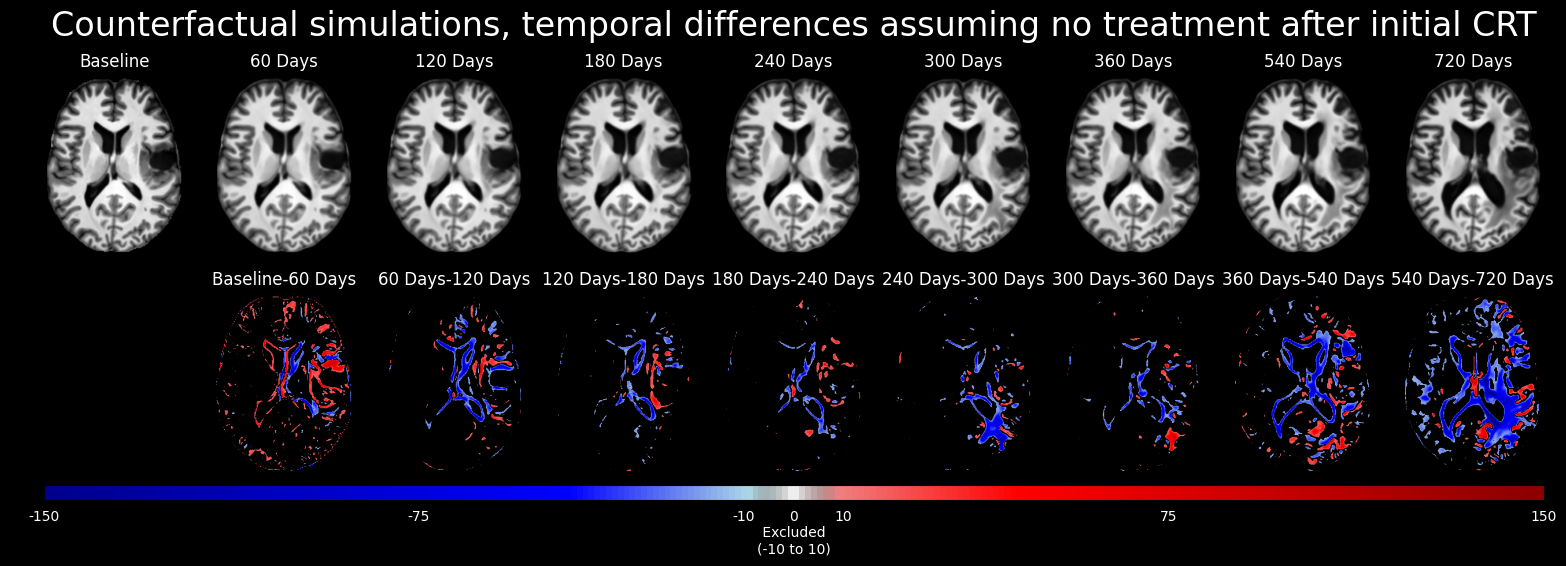}}
    \centering
    \caption{Eight counterfactual generations for every 60 days, varying only in the temporal conditioning. To highlight tumor and tissue morphological evolution, each image is subtracted from the previous one, therefore a difference map for each interval is created. Blue indicates an intensity decrease from one image to the next, while red indicates an intensity increase. The same baseline MRI slice from Figure \ref{fig:example} is included, and right is on the right. Values between -10 and 10 are excluded for clarity.}
    \label{fig:temporal_counterfactuals}
\end{figure}

Figure \ref{fig:temporal_counterfactuals} visualizes the morphological changes over time, by adjusting the input days after treatment for the model while all other variables remain consistent. For the first 240 days, the model mostly predicts general atrophy. Around day 300, the model starts predicting edema, with especially later timepoints such as 540 and 720 days showing increased ventricle size and areas with edema.

\section{Discussion}

This study proposes a rectified flow model for real-time prediction of post-RT MRI for patients with glioma. Our model achieved 0.3 second predictions, with a performance comparable to state-of-the-art models and 250 times fewer sampling steps compared to DDPM. It generates high-quality MRIs for any time point during and after the assumed treatment in just four inference steps based on pre-RT MRIs, RT dose maps, chemotherapy type, and temporal information. SSIM and PSNR are similar to other state-of-the-art models such as TADiff \cite{TADiff}, but unlike TADiff, the model does not require follow-up MRIs during inference and uses 150 times fewer sampling steps. This enables the model to predict follow-up MRIs before treatment is applied and, by changing the treatment-related inputs, to also predict how the MRI would look under alternative treatment choices, enabling counterfactual simulations.

The DSC of tissue and CSF indicate that the real and predicted follow-ups highly overlap, verifying a realistic prediction. However, CSF is systematically overpredicted, especially in later follow-ups. The tissue predictions are more accurate, which may indicate variance in the background segmentation, with some background pixels predicted as CSF. Longitudinally, there is less variance in the predicted follow-ups than the real follow-ups, which is expected since there is no variance in the MRI sequence settings. Patterns in the Jacobian determinants were similar between the real follow-up/baseline and the generated follow-up/baseline, implying that the model's predicted changes are similar to the real changes in the patient’s brain.

Model performance is barely affected by changes to the input conditioning setup, with SSIM, PSNR and MSE showing little difference across the four models. While SSIM is highest in the model without any treatment conditioning, and PSNR and MSE are highest in the model with only chemotherapy conditioning, no conclusions can be made on the effect of treatment conditioning on model performance due to the minor performance gap. This gap could be caused by the complexity of the input information, as well as by variation in the validation or test set. The negligible performance difference results in the model with both RT and chemotherapy conditioning being the optimal model choice, as it is the only model that can simulate counterfactuals for both RT and chemotherapy treatments.

In principle, the predicted follow-up MRIs can be evaluated with the same quantitative pipelines and clinical assessment frameworks used for real follow-up MRIs, such as automated or manual segmentations or the RANO criteria. This enables both automated and expert-based interpretation of the anticipated post-treatment imaging, and allows derivation of trajectories of tumor and tissue response over time, e.g. the evolution of atrophy, edema and cavity changes over months after RT. Such temporally resolved predictions could help differentiate phenomena with distinct time courses, such as early pseudoprogression, later radionecrosis and true tumor progression, by providing an expected baseline evolution for a given treatment. These evaluations should also be extended to quality-of-life applications (QoL), since QoL or similar patient-reported outcomes can be incorporated directly into the training framework and predicted alongside imaging. However, such measurements are not part of standard clinical follow-up in glioma and are mostly collected in the context of clinical trials, and are therefore rarely available in routine datasets. An alternative is to use MRI-based surrogate measures, such as brain-age–like indices, which summarize disease and treatment-related tissue changes into patient-level markers of brain aging and approximate functional reserve \cite{BrainAGEmortality,BrainAGEbiomarker}. Such surrogates could provide QoL–related targets against which both real and generated follow-up MRIs can be evaluated.

The counterfactual generations show that the model responds to RT dose, chemotherapy and time since treatment, which is an important finding indicating the model's reliability despite the small dataset. In general, tissue volume decreases as RT dose increases, and edema can be predicted from a change in systemic treatment. In each of the generated counterfactual MRI with temporal differences, the changes align with previously observed tissue loss after RT \cite{volumeloss,corticalthickness,greymattervolumeloss}. It is important to note that the current simulations are generated solely for demonstration purposes to validate the model's responsiveness to input variables. However, the model is flexible, capable of simulating any realistic treatment variation derived from real clinical considerations.

A key conceptual aspect of this work is that the counterfactuals are explicitly tied to modifiable treatment variables, rather than describing a fixed, non-interventional trajectory of a disease or normal condition. In contrast to similar generative approaches in healthy aging \cite{healthypred} and Alzheimer’s disease \cite{longalz}, where the trajectory is largely non-modifiable and used to visualize or predict natural history, radiotherapy is intrinsically intervention-driven: here, counterfactuals explicitly encode alternative RT or systemic therapy choices that can still be changed before treatment is delivered. Within the same patient, established or hypothetical treatment policies can be explored by perturbing dose, fractionation, and systemic therapy inputs. For instance, in future work, different dose-escalation or hypofractionation \cite{GOLD} schedules could be simulated by altering the temporal and dose conditioning. Alternative Organs-at-risk (OARs) policies can be incorporated directly by adjusting the spatial dose distribution around critical structures and visualizing how different trade-offs between target coverage and OAR sparing affect tumor and tissue over time. This includes known sparing strategies, such as hippocampal avoidance in whole-brain RT \cite{Hippocampalavoidance}, as well as unexplored policies such as thalamic avoidance, that can be investigated in silico. Similarly, incorporating biologically defined target volumes to refine radiation margins from modalities such as FET-PET \cite{FETPET} or advanced quantitative MRI, already used to guide supramarginal resection in neurosurgery \cite{FETPETcomparison}, would allow the model to visualize the impact of various target margins on the surrounding brain. The same conditioning mechanism can be extended to different systemic therapy regimens and their timing relative to RT. Taken together, these capabilities provide a concrete mechanism for running virtual clinical trials in silico, enabling the comparison of established and hypothetical treatment policies before patients are exposed to them in the real world \cite{virtualtrials}.

Although the model's performance is promising, there are limitations. Predictions of the tumor-affected area can be inaccurate, especially when the time point is relatively far from baseline. In particular, clear out-of-distribution situations, such as surgical resection after RT or treatment schemes far outside the training distribution, can lead to implausible predictions. Additionally, non local effects not present in the predicted slice cannot currently be accounted for due to the 2D nature of the model. As the primary goal was to demonstrate the feasibility of rectified flow technology for longitudinal prediction, the current cohort is sufficient to establish the method’s potential. However, scaling this approach for broad clinical application will require a larger, well-processed dataset and a fully 3D model to capture the full variance of clinical scenarios and to enable standard volumetric validation. Furthermore, because quantifying predictive inadequacies with standard metrics is challenging, this future scaling must be accompanied by rigorous clinical validation to ensure that physicians make the same decisions based on predicted images as they would based on real MRIs. Beyond these structural limitations, the current generator operates without explicit constraints of downstream clinical endpoints, such as overall survival. Consequently, there is no built-in upper limit on the plausibility of long-term outcomes; future work could couple the image generative model with survival prediction models to penalize image generation outside the expected survival timeframe \cite{survivalpred}. Finally, future extensions must estimate voxel-wise uncertainty and evaluate how well predicted confidence reflects actual reliability \cite{uncertainty, uncertainty2, uncertainty3}.

While the current technology is not yet mature, successful validation of the model with respect to semantic, quantitative and clinical accuracy could have direct implications for personalized glioma care. Such a tool could help clinicians better balance between tumor control and long-term toxicity such as when a low risk of early recurrence is predicted, a longer interval between follow-up MRI could be used to reduce financial and logistical burden for patients and providers. Having a plausible, individualized picture of expected post-treatment anatomy can also reduce uncertainty for patients and clinicians by making future scenarios more concrete, supporting clearer communication and more informed shared decision making \cite{SDM}. Because the model is agnostic to body region and input channels, the same framework could, in principle, be adapted to other applications and diseases involving RT or even pharmaceutical interventions, where drug type, schedule, and dose play a role analogous to RT dose and fractionation in shaping imaging outcomes. 

\section{Conclusion}

This study presented a novel approach for the generation of counterfactual brain MRI after RT treatment for patients with high-grade glioma. Rectified flow diffusion models could provide a novel method of approaching care for intracranial tumors by predicting morphological changes after treatment. These counterfactual simulations could aid clinicians with creating personalized treatment plans which maximize efficacy and safety of treatment. 

\renewcommand{\baselinestretch}{1} 
\clearpage
\printbibliography

\end{document}